\documentclass[pra,aps,twocolumn]{revtex4}

\usepackage{graphicx}

\usepackage{bm}
\usepackage{amsmath}
\usepackage{amsfonts}
\usepackage{amssymb}
\usepackage{color}
\usepackage{dsfont}

\begin{document}

\title{Entanglement detection by Bragg scattering}

\author{Chiara Macchiavello$^1$ and Giovanna Morigi$^2$}
\affiliation{$^1$ Dipartimento di Fisica and INFN-Sezione di Pavia, Via Bassi 6, 27100 Pavia, Italy\\
$^2$ Theoretical Physics, Saarland University, 66123 Saarbruecken, Germany}

\begin{abstract}
We show how to measure the structural witnesses proposed in  
[P. Krammer {\it et al.}, Phys. Rev. Lett. {\bf 103}, 100502 (2009)] 
for detecting entanglement in a spin chain using photon scattering. 
The procedure, moreover, allows one to measure the two-point correlation function of the spin array. This proposal could be performed in existing experimental platforms realizing ion chains in Paul traps or atomic arrays in optical lattices.
\end{abstract}

\date{\today}

\maketitle


Multipartite entanglement plays a crucial role in various tasks of quantum information, such as in multi-party quantum secret sharing \cite{multi-app} and in the computational speed-up in quantum algorithms \cite{DL,ent-algo}. This motivates the great interest in detecting entanglement in many-body systems. An experimentally feasible method to achieve this task for spin systems was recently proposed in Ref. \cite{Chiara_PRL2010} and later tested in a quantum optical experiment \cite{dicke-exp}. Such a method is based on the construction of entanglement witnesses related to structure factors operators and it is suitable to detect entanglement for various families of multipartite entangled states, such as the Dicke  states, without the need of any {\it a priori} knowledge on the Hamiltonian describing the physical system under consideration. In this paper we discuss specifically how this method can be implemented in ion chains or atomic arrays in optical lattices. In addition, we propose how to measure all elements of the structure factor matrix, thereby allowing one to reconstruct the two-body density matrix of the spin system. 

We first briefly review the structural witnesses method and define the structure factor matrix. Consider a chain of $N$ spin $1/2$ particles which are ordered in a  one-dimensional array at the positions $r_i$ ($i=1,\ldots N$). The structural witness method is based on the measurement of the $3\times 3$ matrix $S({\bf q})$, which is a function of the transfered wave vector ${\bf q}$ and whose elements are given by the static structure factors
\begin{equation}
\label{S:form:factor}
S^{\alpha\beta}({\bf q})
=\sum_{i<j}{\rm e}^{{\rm i}{\bf q}\cdot ({\bf r_i}-{\bf r_j})}\langle S_i^\alpha S_j^\beta\rangle\,.
\end{equation}
In the above expression $S_i^\alpha$ is the $\alpha$ component of the spin 
operator of particle $i$ at position ${\bf r_i}$ ($\alpha=x,y,z$) and the 
average is taken over the initial state of the $N$ spins, whose entanglement 
is to be detected. As shown in Ref. \cite{Chiara_PRL2010}, for different 
values of ${\bf q}$ and different linear combinations of $S^{\alpha\beta}({\bf q})$ it is possible to detect different families of multipartite entangled states. For example, with the choice ${\bf q}=0$ one can construct the witness operator $W_D$
\begin{equation}
\label{S:lin}
\hat{W}_D = \mathds{1} - \frac{2}{N(N-1)}(\hat{S}^{xx}(0)+\hat{S}^{yy}(0)-\hat{S}^{zz}(0))\,,
\end{equation}
where $\mathds{1}$ denotes the identity operator for the $N$-qubit system and 
$\hat S^{\alpha\beta}({\bf q})=\sum_{i<j}{\rm e}^{{\rm i}{\bf q}\cdot 
({\bf r_i}-{\bf r_j})}S_i^\alpha S_j^\beta$ 
are the structure factor operators whose expectation values are given in Eq. 
(\ref{S:form:factor}). Whenever operator $\hat{W}_D$ gives a negative 
expectation value on a density matrix of the spin array $\rho_0$, 
${\rm Tr}\{\hat{W}_D\rho_0\}<0$, we are guaranteed that the $N$-spin state 
under examination is entangled. 
The witness in Eq. \eqref{S:lin} is suitable for detecting symmetric Dicke 
states with any number of excitations, namely its average value for symmetric 
Dicke states is always negative. 

More generally, the structural witness operators take the form
\begin{equation}
\label{W:gen}
\hat{W} = \mathds{1} - \sum_{\alpha=x,y,z} c_\alpha \hat{C}^{\alpha}({\bf q}^\alpha)\;,
\end{equation}
where $c_\alpha$ are real coefficients with $|c_\alpha|\leq 1$, 
$\hat{C}^{\alpha}({\bf q}^\alpha)=\frac{1}{N(N-1)}(\hat{S}^{\alpha\alpha}({\bf q}^\alpha)+\hat{S}^{\alpha\alpha}(-{\bf q}^\alpha))$ and ${\bf q}^\alpha$ denote three possibly different values for the transfered wave vector \cite{dicke-exp}. 

\begin{figure}[hbt]
\begin{center}
\includegraphics[width=0.5\textwidth]{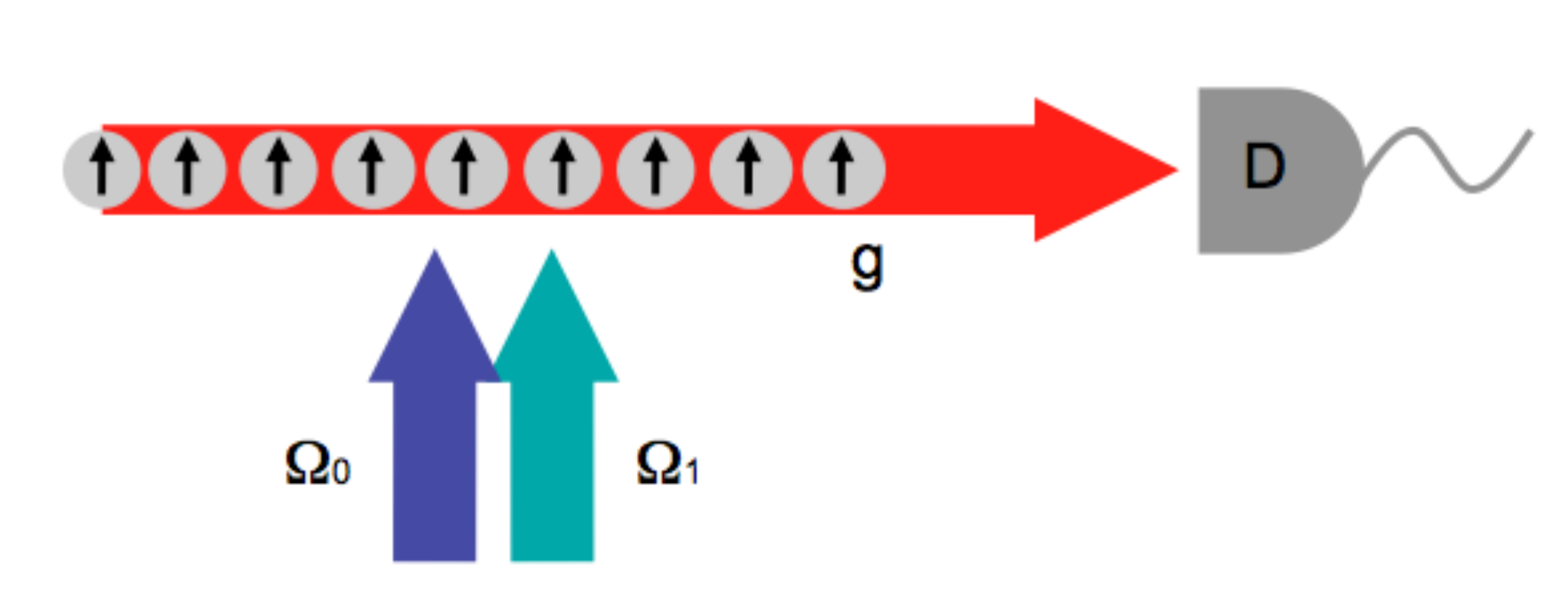}
\caption{\label{Fig1a} (Color online) An entanglement witness for a spin array can be constructed by means of a pump-probe experiment. The intensity of the scattered light allows one to determine the expectation value of the witness operator. The collection efficiency can be increased when the probe field is the mode of a high-finesse resonator \cite{Mekhov}.}
\end{center}
\end{figure}

The underlying detection method relies on the outcome of scattering 
experiments and is therefore particularly suited to detect entanglement in 
many-body systems where local addressing of the individual 
constituent particles 
is not available. The purpose of this paper is to provide a specific 
implementation scheme, 
suggesting a possible pump-probe experiment one can perform on an atomic array or on an ion chain. The setup is sketched in Fig. \ref{Fig1a}. We show that the measurement of the intensity of the probe light allows one to extract all quantities needed for  determining the entanglement witnesses constructed through linear 
superpositions of the quantities in Eq. \eqref{S:form:factor}, as in the
form (\ref{W:gen}). 
In addition, it also allows one to determine all elements of the structure form matrix, therefore measuring the two-point correlation function of the spin system and hence giving access to the two-body density matrix.

Let us first denote by $|0\rangle_j$ and $|1\rangle_j$ the two states of the spin $j=1,\ldots,N$ at position ${\bf r_j}=jd\,\hat{x}$, where $d$ is the interparticle distance and the array is along the $\hat x$-axis. The spins are prepared in the state $\rho_0$ and the elements of Eq. \eqref{S:form:factor} are evaluated over $\rho_0$. These elements can be found by measuring the intensity of a weak probe after implementing a dynamics governed by Hamiltonian \begin{eqnarray}
H_{\rm eff}=\hbar\sum_j\varrho(t)\left(\alpha_0{\rm e}^{{\rm i}\phi}S_j^\dagger+\alpha_1{\rm e}^{-{\rm i}\phi}S_j\right){\rm e}^{{\rm i}{\bf q}\cdot {\bf r_j}}a+{\rm H.c.}\,,\label{H:eff}
\end{eqnarray}
where $S_j^\dagger=|1\rangle_j\langle0|$, $S_j=|0\rangle_j\langle 1|$, such that $S_j^x=S_j^{\dagger}+S_j$ and $S_j^y=-{\rm i}(S_j^{\dagger}-S_j)$. The parameters $\alpha_0,\alpha_1$ are real-valued frequencies, $\phi$ is a phase, ${\bf q}$ is the difference between probe and pump wave vectors, and $a$ the annihilation operator of the probe field. The dimensionless function $\varrho(t)$ gives the temporal form of the excitation, which warrants that the dynamics is in the linear response regime, so that the scattered light depends on the initial state. Here, $\varrho(t)=0$ for $t<0$ and ${\rm max}|\varrho(t)|=1$. Moreover, the pulse temporal duration is finite and the variation is sufficiently smooth to warrant the effective dynamics described in Eq. \eqref{H:eff}. In the following we will characterize its characteristic temporal duration by $\Delta t$.

Since the number of scattered photons at each experimental run must be small, a high collection efficiency of the scattered light is required. This can be achieved when the probe is the mode of a high-finesse resonator \cite{Mekhov} and could be realized in existing experimental platforms, where atomic arrays \cite{Zimmerman,Vuletic} or ion crystals \cite{Drewsen} have been confined and strongly coupled with a cavity field. In the following we shall consider that the probe field is one mode of a high-finesse ring cavity, which strongly couples with an optical dipole transition.

\begin{figure}[hbt]
\begin{center}
\includegraphics[width=0.25\textwidth]{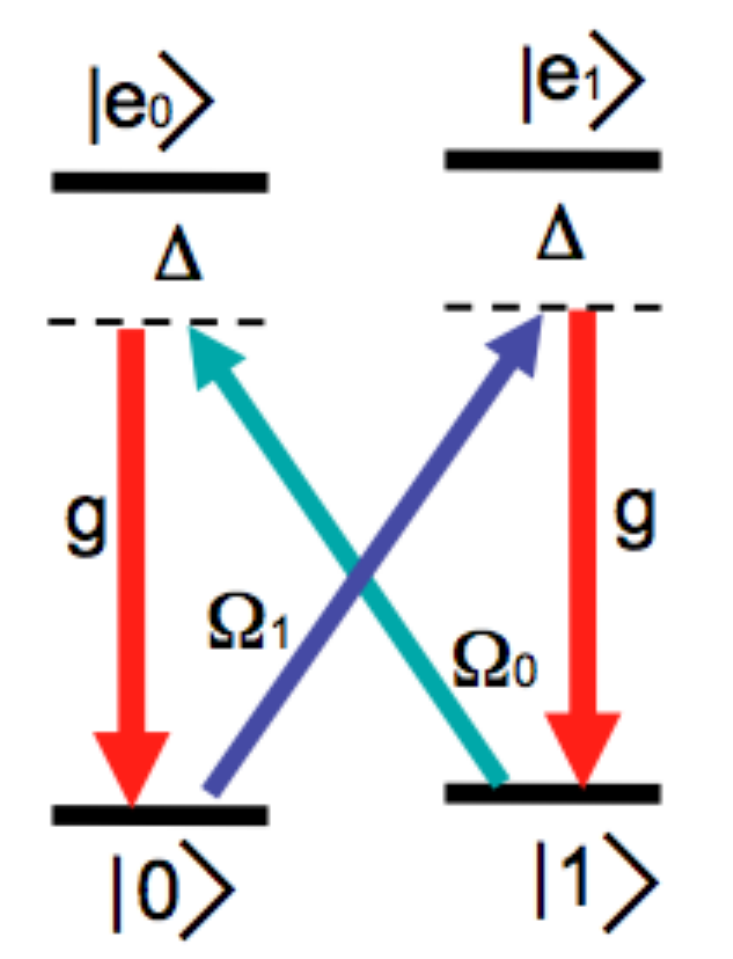}
\caption{\label{Fig1b} (Color online) A possible excitation scheme, where two hyperfine states of the ground state multiplet are the spin states. In the scheme here discussed, the realization of the witness lies on the capability to adjust the amplitude and the relative phase of the lasers with Rabi frequency $\Omega_0$ and $\Omega_1$.}
\end{center}
\end{figure}

In order to give a detailed derivation of Hamiltonian \eqref{H:eff}, we first specify the atomic states composing the spin transition of the atoms composing the array. We assume that the spin of a single atom is composed by radiatively stable states of the hyperfine multiplet of state $S_{1/2}$. This example is taken for convenience, since the treatment in the following can be extended to other pairs of radiatively stable states. We denote the spin states by $|0\rangle\equiv |S_{1/2},-1/2\rangle$ and  $|1\rangle\equiv |S_{1/2},+1/2\rangle$, while the excited states (which will be needed in order to obtain the effective Hamiltonian) are  $|e_0\rangle\equiv |P_{1/2},-1/2\rangle$ and   $|e_1\rangle\equiv |P_{1/2},+1/2\rangle$. For simplicity, we  assume that the states $|0\rangle,|1\rangle$ are degenerate and their energy  is set to zero, while states  $|e_0\rangle, |e_1\rangle$ are both at frequency $\omega_0$, so that the Hamiltonian for the relevant electronic degrees of freedom reads $H_{\rm el}=\hbar\omega_0\sum_j\left(|e_0\rangle_j\langle e_0|+|e_1\rangle_j\langle e_1|\right)$. The electronic levels are coupled by cavity field and lasers as shown in Fig. \ref{Fig1b}. The pump lasers, in particular, have wave vector ${\bf k_L}$, frequency $\omega_L$, and couple transitions $|1\rangle_j\to |e_0\rangle_j$ and $|0\rangle_j\to |e_1\rangle_j$ with maximum Rabi frequency $\Omega_0$ and $\Omega_1$, respectively. The cavity mode, which acts as a probe, is linearly polarized and at frequency $\omega_c$ and wave vector ${\bf k}$. We denote by $a_1$ and $a_1^\dagger$ the annihilation and creation operators of a cavity-mode photon, and by $a_2$ and $a_2^\dagger$ the corresponding operators for the second, degenerate cavity mode at wave vector $-{\bf k}$. Both modes drive transitions $|0\rangle_j\to|e_0\rangle_j$ and $|1\rangle_j\to|e_1\rangle_j$ with vacuum-Rabi frequency $g$. The Hamiltonian governing the coherent dynamics of spins and cavity modes reads 
\begin{eqnarray}
H_L&=&-\hbar \delta_c \left(a_1^{\dagger}a_1+a_2^{\dagger}a_2\right)-\hbar\Delta\sum_j\left(|e_0\rangle_j\langle e_0|+|e_1\rangle_j\langle e_1|\right)\nonumber\\
&&+\hbar g\sum_j\left[\left(|e_0\rangle_j\langle 0|+|e_1\rangle_j\langle 1|\right)a_1{\rm e}^{{\rm i}{\bf k}\cdot {\bf r_j}}+{\rm H.c.}\right]\nonumber\\
&&+\hbar g\sum_j\left[\left(|e_0\rangle_j\langle 0|+|e_1\rangle_j\langle 1|\right)a_2{\rm e}^{-{\rm i}{\bf k}\cdot {\bf r_j}}+{\rm H.c.}\right]\nonumber\\
&&+\hbar\Omega_0 \varrho(t)\sum_j\left[|e_0\rangle_j\langle 1|{\rm e}^{{\rm i}{\bf k_{L}}\cdot {\bf r_j}}{\rm e}^{-{\rm  i}\phi}+{\rm H.c.}\right]\nonumber\\
&&+\hbar\Omega_1 \varrho(t)\sum_j\left[|e_1\rangle_j\langle 0|{\rm e}^{{\rm i}{\bf k_{L}}\cdot {\bf r_j}}{\rm e}^{{\rm  i}\phi}+{\rm H.c.}\right]
\end{eqnarray}
and is reported in the reference frame rotating at the carrier frequency $\omega_L$ of the laser pulses, with $\Delta=\omega_L-\omega_0$ and $\delta_c=\omega_L-\omega_c$, respectively. The phase $\phi$ determines the relative phase between the fields, since we assumed that $\Omega_0,\Omega_1,g$ are real-valued. The incoherent dynamics due to the radiative decay of the excited state at rate $\gamma$ can be suppressed by driving far-off resonant Raman transitions, which corresponds to choosing values of the detuning  such that $|\Delta|\gg g,\Omega_j,|\delta_c|,\kappa,\gamma$, where $\kappa$ is the linewidth of the cavity modes. In this limit the atomic state follows adiabatically the cavity field, namely 
\begin{eqnarray}
|0\rangle_j\langle e_0|&=&\frac{g}{\Delta}|0\rangle_j\langle 0|\left(a_1{\rm e}^{{\rm i}{\bf k}\cdot {\bf r_j}}+a_2{\rm e}^{-{\rm i}{\bf k}\cdot {\bf r_j}}\right)\\
& &+\frac{\Omega_0}{\Delta}\varrho(t)|0\rangle_j\langle 1|{\rm e}^{-{\rm i}\phi}{\rm e}^{{\rm i}{\bf k_L}\cdot {\bf r_j}}\nonumber\\
|1\rangle_j\langle e_1|&=&\frac{g}{\Delta}|1\rangle_j\langle 1|\left(a_1{\rm e}^{{\rm i}{\bf k}\cdot {\bf r_j}}+a_2{\rm e}^{-{\rm i}{\bf k}\cdot {\bf r_j}}\right)\\
& &+\frac{\Omega_1}{\Delta}\varrho(t)|1\rangle_j\langle 0|{\rm e}^{{\rm i}\phi}
{\rm e}^{{\rm i}{\bf k_L}\cdot {\bf r_j}}\,.\nonumber
\end{eqnarray}
These equations hold at the lowest non-vanishing order in the perturbative expansion in $1/|\Delta|$ and when the width of the pulse is such that $\Delta t\gg 1/|\Delta|$. These values are substituted in the Heisenberg-Langevin equation for the cavity mode, $\dot{a}_\ell=-\kappa a_\ell +\sqrt{2\kappa}a_{\ell,\rm in}+[a_\ell,H_L]/({\rm i}\hbar)$, where $a_{\ell,\rm in}(t)$ is the input noise field, 
satisfying the relations $\langle a_{\ell,\rm in}(t)\rangle=0$, 
$\langle a_{\ell,\rm in}(t)a_{k,\rm in}^\dagger(t')\rangle=\delta_{k\ell}
\delta(t-t')$ ($j,k=1,2$) \cite{Walls}. Setting $a\equiv a_1$, in the linear response regime the solution reads
\begin{eqnarray}
\label{eq:a}
&&a\simeq {\rm e}^{-(\kappa-{\rm i}\delta_c') t} a(0)+\sqrt{2\kappa}\int_0^t{\rm d}t'{\rm e}^{-(\kappa-{\rm i}\delta_c')  (t-t')}a_{\rm in}(t')\nonumber\\
&&-{\rm i}\frac{g^2}{\Delta}\sum_{j}{\rm e}^{-{\rm i}2{\bf k}\cdot {\bf r_j}}\left(\frac{1-{\rm e}^{-(\kappa-{\rm i}\delta_c')t}}{\kappa-{\rm i}\delta_c'}\right)a_2(0)
\nonumber\\
&&-f(t)\sum_j\left(\alpha_0{\rm e}^{{\rm i}\phi}S_j(0)+\alpha_1{\rm e}^{-{\rm i}\phi}S_j^{\dagger}(0)\right){\rm e}^{{\rm i}{\bf q_1}\cdot {\bf r_j}}
\end{eqnarray}
where ${\bf q_1}={\bf k_L}-{\bf k}$, $\alpha_0=g\Omega_0/\Delta$,  $\alpha_1=g\Omega_1/\Delta$, and $f(t)={\rm i}\int_0^t{\rm d}\tau {\rm e}^{-(\kappa-{\rm i}\delta_c')(t-\tau)}\varrho(\tau)$. Note that the detuning $\delta_c'=\delta_c+g^2/\Delta$ includes the dynamical Stark shift of the spin states \cite{Rist}. Equation \eqref{eq:a} has been obtained assuming $\kappa\gg |\alpha_\ell|$, so that the photons scattered by the atoms into the cavity are not reabsorbed but emitted by the cavity. The last line of the right-hand side of this equation, in particular, corresponds to the dynamics governed by the target Hamiltonian, Eq. \eqref{H:eff}. For the considered setup, this is the only relevant term  which determines the intensity of the probe field, as we will show. 

The intensity of the probe field at the cavity output  reads $I_{\rm out}=\langle a_{\rm out}(t)^{\dagger}a_{\rm out}(t)\rangle$, where $a_{\rm out}(t)=\sqrt{2\kappa}a(t)+a_{\rm in}(t)$. It is evaluated using Eq. \eqref{eq:a}, the properties of the input operators, and assuming that the cavity modes are initially in the vacuum field and can be cast in the form $I_{\rm out}=2\kappa |f(t)|^2 \left( I_0+I_{\rm int}\right)$, where $I_0$ is the sum of the intensities scattered by each atom, while $I_{\rm int}$ is the intensity due to the interfence between the scattered waves at different sites. In detail, 
\begin{equation}
\label{I:0}
I_0=N\left[\left(|\alpha_x|^2+|\alpha_y|^2\right)+2{\rm Im}\{\alpha_x\alpha_y^*\}\sum_{k=1}^N\langle S_k^z\rangle/N\right]\,,
\end{equation}
and 
\begin{eqnarray}
\label{I:int}
I_{\rm int}&=&\sum_{k\neq \ell}{\rm e}^{-{\rm i}{\bf q_1}\cdot ({\bf r_k}-{\bf r_\ell})}\left(|\alpha_x|^2\langle S_k^xS_\ell^x\rangle+|\alpha_y|^2\langle S_k^yS_\ell^y\rangle\right.\nonumber\\
&&\left.+\alpha_x^*\alpha_y\langle S_k^xS_\ell^y\rangle+\alpha_x\alpha_y^*\langle S_k^yS_\ell^x\rangle\right)\,.
\end{eqnarray}
The coefficients depend on the lasers amplitudes and relative phase as follows 
\begin{eqnarray}
&&\alpha_x=\frac{\alpha_1{\rm e}^{{\rm i}\phi}+\alpha_0{\rm e}^{-{\rm i}\phi}}{2}\\
&&\alpha_y=\frac{\alpha_1{\rm e}^{{\rm i}\phi}-\alpha_0{\rm e}^{-{\rm i}\phi}}{2{\rm i}}\,.
\end{eqnarray}

The possibility to vary these coefficients gives access to the general 
structural witnesses (\ref{W:gen}), as can be 
easily shown using the identity
\begin{equation}
\sum_{k\neq \ell}{\rm e}^{-{\rm i}{\bf q}\cdot ({\bf r_k}-{\bf r_\ell})}\langle S_k^\alpha S_\ell^\beta\rangle=S^{\alpha\beta}({\bf q})+S^{\alpha\beta}(-{\bf q})\,.
\end{equation}
For instance, the element $\hat C^{x}({\bf q})$ is found by setting 
$\alpha_y=0$ 
and $\alpha_x=\alpha$, which corresponds to choosing $\Omega_0=\Omega_1$ and 
$\phi=0$. 

This setup allows also to measure the off-diagonal components in the matrix 
(\ref{S:form:factor}). For example, by performing the measurement for different values of $\Omega_0,\Omega_1,\phi$ and on the intensity of the second cavity mode allows one to determine the components  $S^{xy}({\bf q})$ and  $S^{yx}({\bf q})$. The other elements involving also the $z$ component,  
namely $S^{zy}({\bf q})$ and  $S^{yz}({\bf q})$ ($S^{zx}({\bf q})$ and  
$S^{xz}({\bf q})$), are found by performing the same 
measurement procedure after the Hadamard transformation 
$\mathcal U_H^{y}=\otimes_j(S_j^x+S_j^z)/\sqrt{2}$ 
($\mathcal U_H^{x}=\otimes_j(S_j^y+S_y^z)$) has been made on the qubits of the initial state: the density matrix over which one performs the trace in Eqs. 
\eqref{I:0} and \eqref{I:int} is now 
$\rho'=\mathcal U_H^{\alpha}\rho_0 \mathcal U_H^{\alpha}$, which corresponds to a Raman pulse composed by classical fields \cite{Ozeri}. 
The net effect is hence to transform the operators $S_j^x$ into $S_j^z$ 
($S_j^y$ into $S_j^z$) and vice versa, so that, for example, 
${\rm Tr}\{S_k^xS_\ell^x\rho'\}= {\rm Tr}\{S_k^zS_\ell^z\rho_0\}$. 
In this way also the term
$\hat C^{z}({\bf q})$ of the general structural witness can be accessed, as well as all other elements needed in order to reconstruct the two-body density matrix. 

These results show that in the case of Dicke states the entanglement witness (\ref{S:lin}) can be determined by controlling the amplitudes and phases of the driving laser. One possibility for extracting the value at ${\bf q}={\bf 0}$ is when the array is along the cavity axis and orthogonal to the laser direction, setting $d\hat{x}\cdot {\bf k}=2\pi n$, where $n$ is an integer and we assumed that the chain is along the $x$ axis. We note that localization about the minima is required, so that there is no hopping from site to site. The effect of the vibrations about the minima will tend in general to lower the interferometric contrast, however in the Lamb-Dicke regime this detrimental effect is small \cite{Eschner}. 

In summary, we have presented a specific setup to implement a recently
proposed multipartite entanglement detection method \cite{Chiara_PRL2010} in a chain of atoms or ions in an optical lattice. The example we provide here shows explicitly how the measurement could be performed with current-day technology \cite{Zimmerman,Drewsen,Vuletic}. These calculations can be extended to other setups based on the atomic microscopes in Ref. \cite{Bloch}, or on quantum non-demolition measurement like the ones discussed in Ref. \cite{Hammerer}.  
We mention that recently an entanglement detection method based on the 
proposal \cite{Chiara_PRL2010} and on a similar protocol \cite{Cramer} has 
been experimentally realized for the motional degrees of freedom of cold 
atoms in optical lattices \cite{LENS}.

The present entanglement detection scheme is based on the measurement of two-point correlation functions, and provides, amongst others, a determination of the magnetic susceptibility, which at thermodynamic equilibrium and for specific Hamiltonians constitute a macroscopic entanglement witness \cite{vlatko}. We conclude by pointing out that measurements of higher order correlation functions of the scattered field in the proposed system allows one to determine higher order correlation functions of the spin operators \cite{Mintert}, which are at the basis of witnesses, for instance, of GHZ-like multipartite states \cite{w-jmo}.


We acknowledge support by the EU (IP AQUTE, STREP PICC) and the German Research Foundation (DFG). The authors thank D. Bru\ss, D. Lucas, C. Townsend, and P. Verrucchi for stimulating discussions and hospitality by the Clarendon Laboratory at the University of Oxford.

\end{document}